\documentclass[aps,twocolumn,showpacs]{revtex4}
\usepackage{amssymb}
\usepackage{amsmath}
\usepackage{epsfig}

\newcommand{\cn}{\rm cn}

\newcommand{\dn}{\rm dn}
\begin{document}
\title{\bf Bistable light detectors
with nonlinear waveguide arrays}
\author { R. Khomeriki${}^{1,2}$,  J. Leon${}^1$}
\affiliation {(${\ }^1$) Laboratoire de Physique Th\'eorique et 
Astroparticules\\
CNRS-UMR5207, Universit\'e Montpellier 2, 34095 Montpellier  (France)\\
 (${\ }^2$) Physics Department, Tbilisi State University,
0128 Tbilisi (Georgia)}
\begin{abstract} Bistability induced by nonlinear Kerr effect in arrays of
coupled waveguides is studied and shown to be a means to conceive light
detectors that switch under excitation by a weak signal. The detector is
obtained by coupling two single 1D waveguide to an array of coupled waveguides
with adjusted indices and coupling. The process is understood by analytical
description in the conservative and continuous case and illustrated by
numerical simulations of the model with attenuation.\end{abstract}
\maketitle

\paragraph*{Introduction.}

Arrays of adjacent optical dielectric waveguides coupled by power exchange
between guided modes \cite{yariv} \cite{kivsh-agra}, have allowed to conceive
devices possessing extremely rich optical properties, already at the linear
level, such as anomalous refraction and diffraction \cite{pertsch}. 

At nonlinear level, for  intensity-dependent refractive index (optical Kerr
effect), these waveguide arrays become soliton generators \cite{christ-joseph},
as experimentally demonstrated in
\cite{eisenberg,morandotti,mandelik,expe1,yuri,fleisher1,fleisher2}. The model is a discrete
nonlinear Schr\"odinger equation (NLS) and  nonlinearity then manifests by
self-modulation of an input signal (injected radiation) that propagates as a NLS
discrete soliton \cite{yuri,andrey}. These systems possess also intrinsically
discrete properties \cite{aceves} and the geometry can be varied to manage
dispersion \cite{mark}.
\begin{figure}[b] \centerline
{\epsfig{file=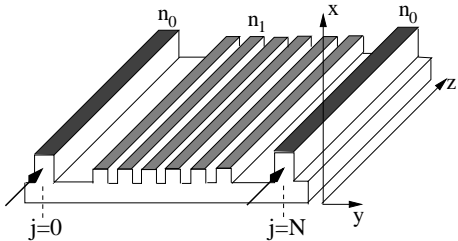,width=0.7\linewidth}} \caption {\it  Waveguide
array (index profile) driven by 2 lateral waveguides of index $n_0>n_1$. The
arrows symbolize the injected radiation. }
\label{fig:device}\end{figure}

A fundamental property of nonlinear systems that has not been considered
in waveguides arrays is the bistability induced by nonlinearity. It is the
purpose of this work to propose and study a device where bistability, and
consequent switching properties, could be observed and used to conceive for
instance a detector of light sensitive to extremely weak signal.

We shall make use of the possibility to drive a waveguide array, in the
forbidden band gap, through directional coupling by boundary waveguides, which
results in the generation of  (discrete) gap solitons \cite{khomeriki}, produced
by nonlinear supratransmission \cite{nst}, and used to conceive resonators with
nonlinear eigenstates \cite{jl}. 

\begin{figure}[ht] \centerline
{\epsfig{file=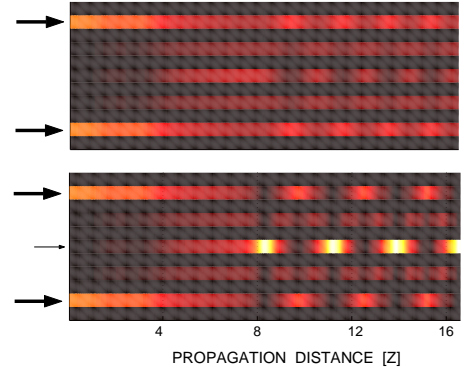,width=0.7\linewidth}} \caption {\it 
(Color inline) Flux intensity along the waveguides driven at the threshold of
bistability. This is the result of numerical simulations of the model 
\eqref{discrete} with $N=4$, $Q=Q_0=4$, $\gamma=\gamma_0=0.001$, $V=0.5$, and
input amplitudes $b=3.0946$ (large arrows).  In the lower plot the small arrow
represents a signal of amplitude $0.001$ enough to trigger the switch.}
\label{fig:map}\end{figure}

The waveguide array of figure \ref{fig:device}, in nonlinear Kerr regime,
driven by two single waveguides of index $n_0>n_1$, is operated in such a way
that the guided modes in the two lateral waveguides would not {\it linearly
propagate} in the array due to index difference. Nonlinear induced bistability
allows then to adjust the input radiation such as to work at the bifurcation
point and switch from a state of vanishing intensity output (from the central
waveguides) to a state of strong intensity output, as displayed on
fig.\ref{fig:map} that represents the intensity of the flux in the array when
operated close to the bifurcation point. As the switch can be operated by a
very weak signal (0.03\% of the input amplitude for fig.\ref{fig:map}), the
device is a candidate for an  {\it ultrasensitive bistable detector}. 

After having briefly recalled the derivation of the discrete NLS model for the
envelope of the guided modes and the proper boundary conditions that results
from the physical context of fig.\ref{fig:device}, we describe theoretically
and numerically the bistability process. The theory calls to the continuous
limit which gives an accurate description  of the discrete case with enough
waveguides and strong coupling.

\paragraph*{The model.}

Assuming instantaneous response of the Kerr medium (of nonlinear susceptibility
$\chi$) and planar wave guides along the vertical direction $x$, the electric
field can be sought as $\vec E(\vec r,t)=\vec u_x( E(y,z)\ e^{-i\omega
t}+c.c.)/\sqrt{3\chi}$ and Maxwell's equations reduce to
\begin{equation}
 (\frac {\partial^2 }{\partial y^2}+
 \frac {\partial^2\ }{\partial z^2})\ E +\frac{\omega^2}{c^2}
\left( n^2+ \mid E\mid^2\right)E=0,
\label{maxwell}\end{equation}
where the linear index $n=n(\vec r)$ may vary in space.
Interested in the stationary regime, the fast oscillating nonlinear term has
been discarded (assuming no phase matching between third harmonics).

We consider the index profile of fig.\ref{fig:device} and assume
a field component $E(y,z)$ in the form
\begin{equation}\label{E}
E(y,z)=\epsilon \sum_{j}{\cal E}_j(\epsilon y)\psi_j(\epsilon^2z)e^{i\beta z},
\end{equation}  
where ${\cal E}_j$ is a guided mode, $\psi_j$ its amplitude and
$\epsilon$ a small parameter to be defined later. The slow
variation along $z$ results from the coupling between adjacent waveguides.
Guided modes in both lateral waveguides, that do not propagate transversaly  in
the array, result from the necessary condition
\begin{equation}\label{condition}
\beta^2-\frac{\omega^2}{c^2}n_0^2<0,\quad
\beta^2-\frac{\omega^2}{c^2}n_1^2>0.\end{equation}
Then, following \cite{ziad}, we insert \eqref{E} in \eqref{maxwell},
integrate over $y$ and obtain the coefficients that couple the driving lateral
wave guides to the array ($C_0$), the array waveguides together ($C_1$), and
that result from the overlapping of modes ${\cal E}_j$ across the separation
between waveguides.  Upon defining the dimensionless variable
$z'=z\,\omega^2/(2\beta c^2)$  the coupling coefficients are scaled by
\begin{equation}
Q=\frac{c^2}{\omega^2}C_1,\quad Q_0=\frac{c^2}{\omega^2}\frac{C_0^2}{C_1}.
\end{equation}
and we define the essential constant
\begin{equation}\label{cns}
V=n_0^2-n_1^2-2Q,\quad V>0,\quad V\sim\epsilon^2.\end{equation}
This defines the small parameter $\epsilon$ used in expansion \eqref{E},
and provides the necessary and sufficient condition ($V>0$) for evanaescent
waves along transverse direction  in the array.

We obtain eventually the model (forgetting the prime on $z$ and setting
$\epsilon=1$ to go back to physical units)
\begin{align} 
&i(\partial_z+\gamma_0)\psi_{0} =Q_0\psi_1,\quad 
i(\partial_z+\gamma_0)\psi_{N}=Q_0\psi_{N-1}\label{discrete} \\
&i(\partial_z+\gamma)\psi_j 
+Q\left(\psi_{j-1}+\psi_{j+1}-2\psi_j\right)+|\psi_j|^2\psi_j=V\psi_j,
 \nonumber\end{align} 
where $j$ runs from $j=1$ to $j=N-1$. Attenuation has been included (imaginary
parts $\gamma$ and $\gamma_0$ of the dielectric constants) to account for a
realistic physical situation, and the actual amplitudes in the lateral driving
waveguide have been rescaled by replacing $C_0\psi_{0,N}$ by $C_1\psi_{0,N}$.

Prescription of the injected energy flux in the lateral waveguides results in
defining the amplitudes $\psi_0$ and $\psi_{N}$ in $z=0$ (say $b$), which are
associated to a set of vanishing $\psi_j(0)$ for the rest of the array. The
initial condition is thus
\begin{equation}\label{init}
\psi_0(0)=\psi_N(0)=b,\quad  \psi_j(0)=0,\quad j=1\cdots N-1\end{equation}
which complete  equations \eqref{discrete}. The point is that nonlinearity
causes the existence of a threshold $b_s$ allowing to take advantage 
of bistabilility of stationary states, as described now analytically
and numerically.

\paragraph*{Analytic description.}

The continuous limit of \eqref{discrete} is obtained for
\begin{equation}
y=j\sqrt{1/Q},\quad \psi_j(z)=\psi(y,z),\end{equation} 
(note that this dimension $y$ is not the physical dimension $y$ of expression
\eqref{E}) and reads (without attenuation)
\begin{equation} \label{drivenNLS}
i\psi_z+\psi_{yy}+|\psi|^2\,\psi=V\psi.\end{equation}
The boundary data that represent the injected flux are taken as constants,
namely
\begin{equation}
\psi(0,z)=\psi(L,z)=b,
\end{equation}
As far as the light tunnels from the driving waveguides to the array, the
amplitude of light in the driving waveguides decays along $z$. However, if the
coupling $Q_0$ and the attenuation $\gamma_0$ are small enough, the 
intensities in the driving waveguides decay {\em slowly} along $z$ and the
above boundary-value problem quite correctly matches numerical simulation of
the discrete model \eqref{discrete}.

The stationary solutions of the above problem are obtained by assuming a real
valued solution depending on $y$ only, namely $\psi(y,z)=\psi(y)$ solution of
\begin{equation}
\psi_{yy}+\psi^3=V\psi, \quad \psi(0)=\psi(L)=b. \label{solution}
\end{equation}
From the symmetry of the above bounds, a uniformly bounded function requires
that there exists $y_0$ such that
\begin{equation}
\psi_y(y_0)=0, \quad \psi(y_0)=B. \label{boundary}
\end{equation}
The values $b$ and $B$ are referred respectively to as input and output 
amplitudes of the field. By integrating \eqref{solution} with
\eqref{boundary}, one gets
\begin{equation}
(\psi_y)^2=\frac{1}{2}(B^2-\psi^2)(\psi^2+B^2-2V). \label{elliptic}
\end{equation}
This equation has solutions of different types \cite{byrd} depending on the
relative values of $B$ and $V$. In particular, an input value $b$ of the
boundary driving can produce different output amplitudes $B$. This
property holds when the input amplitude $b$ is less than a value  $b_s$,
called the supratransmission threshold, above which the boundary driving 
induces an instability which generates soliton emission in the array
\cite{instab}.

Case I: $B^2>2V$. The integration of \eqref{elliptic} on $[0,L]$
produces the relation
\begin{equation}
b=B\cdot \cn[\beta L/2,k], \label{cn}
\end{equation}
where $\cn[\cdot,k]$ is the cosine-amplitude Jacobi elliptic
function of modulus $k$. The parameters $\beta$ and $k$ obey the relation
\begin{equation}
\beta^2=B^2-V, \quad k^2=\frac{B^2}{2(B^2-V)}, \label{I}
\end{equation}
while the whole solution \eqref{solution} has the following form
\begin{equation}
\psi^I(y)=B\cdot\cn\left[\beta(y-L/2),k\right]. \label{solutionI}
\end{equation}
Case II: $V<B^2<2V$. The integration of \eqref{elliptic} gives here
\begin{align}
&b=B\cdot \dn[\beta L/2,k], \label{dn1} \\
&\beta^2=\frac{B^2}{2}, \quad k^2=\frac{2(B^2-V)}{B^2},\label{II}\\
&\psi^{II}(y)=B\cdot\dn\left[\beta(y-L/2),k\right]. \label{solutionII}
\end{align}
Case III: $B^2<V$. In this case we obtain 
\begin{align}
&b=\frac{B}{\dn[\beta L/2,k]}, \label{dn2} \\
&\beta^2=\frac{2V-B^2}{2}, \quad k^2=\frac{2(V-B^2)}{2V-B^2}, \label{III}\\
&\psi^{III}(y)=\frac{B}{\dn\left[\beta(y-L/2),k\right]}. \label{solutionIII}
\end{align}
This last solution has a threshold amplitude $b=b_s$ reached when the function
$b(B)$ in \eqref{dn2} has a vanishing derivative. The related expression of
$b_s$ is not explicit although easily evaluated numerically. Note that in
general one has a finite set of thresholds (their number depend on the lenght
$L$) but only the first one is of interest for our purpose. Note also that the
above three solutions reach their maximum amplitude $B$ in the array center
$y=L/2$. 

\paragraph*{Bistable behavior.}

\begin{figure}[ht] \centerline {\epsfig{file=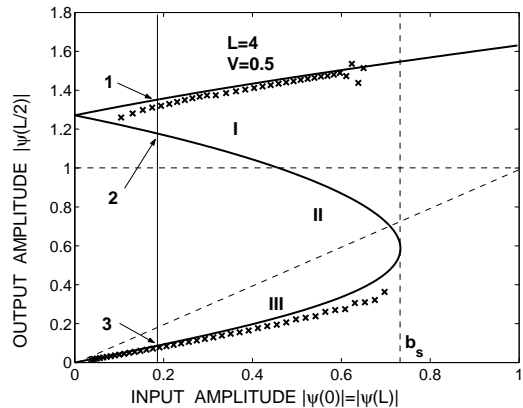,width=0.8\linewidth}}
\caption {\it  Hysteresis loop for the input-output amplitude values obtained
from the analytical solutions of type I \eqref{cn}, II \eqref{dn1} and III
\eqref{dn2} respectively. Crosses are the results of numerical simulations. The
dashed vertical line represents the value of the supratransmission threshold
$b_s$, and ther two dashed lines set upper and lower limits for the appearance of type II solution. Crossing of the full vertical line with the hysteretic curve defines input-output values 1, 2 and 3 chosen for fig.\ref{profiles}.}
\label{schroedhyst}\end{figure}

\begin{figure}[ht] \centerline
{\epsfig{file=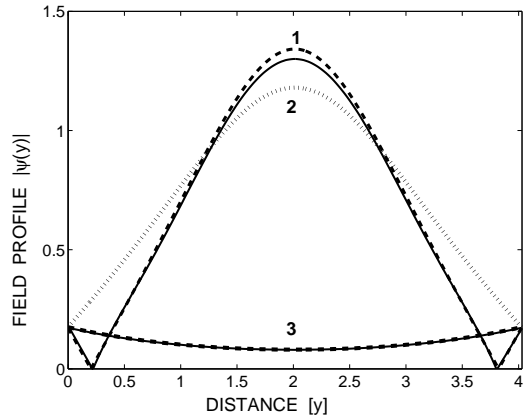,width=0.8\linewidth}}  
\caption {\it  Analytic solutions (dashed lines) corresponding to the single
input amplitude $b=0.18$  compared to numerical experiments (solid lines).}
\label{profiles}\end{figure}

The numerical simulations of system \eqref{discrete}, now compared to the above
analytical solutions, are performed  with $N=18$,  $Q=20$, $Q_0=0.05$ and
$V=0.5$. The system locks to a stationary solution thanks to the attenuation
factors $\gamma=0.02$ and $\gamma_0=0.002$. In order to avoid an initial shock,
the coupling $Q_0$ is actually smoothly set by using $Q_0(1-e^{-z/5})$. The
input-output dependence resulting from of  (\ref{cn}), (\ref{dn1}) and
(\ref{dn2}) is plotted on  fig.\ref{schroedhyst} and the profiles of the
different exact solutions \eqref{solutionI}, \eqref{solutionII} and
\eqref{solutionIII} corresponding to the different output amplitudes are
presented in fig.\ref{profiles}. 


\begin{figure}[t] \centerline
{\epsfig{file=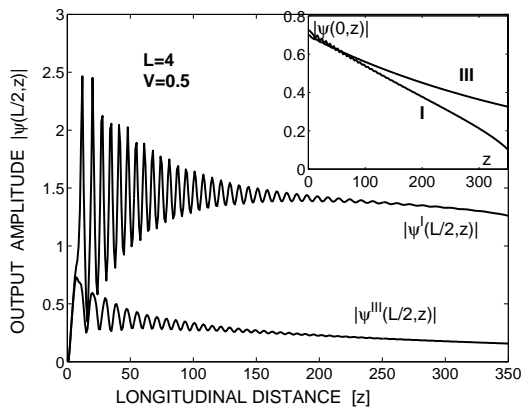,width=0.8\linewidth}}
\caption {\it Evolution of input (inset) and output (main plot) amplitudes
along the propagation direction $z$. Note the difference between output
amplitudes $|\psi(L/2,z)|$ in two regimes although the initial values of the
amplitudes in the boundary waveguide $b$ slightly differ from each
other.}  \label{schroedevol}\end{figure}

The regimes of stationary solutions are reached by numerical simulations of
\eqref{discrete} as follows. First we inject into the driving waveguides the
beams with the initial amplitude \eqref{init} for $b<b_s$. This produces the
type III solution,  a nonlinear analogue to the evanescent wave profile. As the
amplitudes $|\psi_0(z)|$ and $|\psi_N(z)|$ in the driving waveguides decay
along $z$ we monitor the amplitude $|\psi_{N/2}(z)|$ at the middle of the array
as a function of the driving amplitudes. At some distance $z_0$ from the
origin, the system stabilizes to a stationary profile (as shown by
fig.\ref{schroedevol}) and we measure there the corresponding amplitudes 
$B=|\psi_{N/2}(z_0)|$ and $b=|\psi_0(z_0)|$, plotted then as the crosses along
the lower branch (III) in fig.\ref{schroedhyst}.
(Note that only the solutions of  type I and III can be reached by numerical
simulations, while the type II is unstable and eventually decays to the
stable solution III. Note also that longer lengths would allow for
multiple output amplitudes, which we do not consider here.)

The same procedure is used when injecting the beams with $b>b_s$. After  a
regime of gap solitons emission, the amplitudes along the lateral waveguides
decrease below the threshold $b_s$, allowing the system to lock to the solution
of type I. The obtained numerical values are plotted (crosses) along the upper
branch (I) of the hysteresis loop in fig.\ref{schroedhyst}. 

The analytical curves slightly overestimate the numerical results because of
the attenuation (optical losses in the laboratory experiments) included in our
numerical simulations.  Another consequence of the attenuation  is the 
existence of a second (nonzero) threshold input amplitude where the solution
bifurcates back from the regime I to the regime III. For a vanishing
attenuation, this second threshold would be exactly zero according to the
analytical solution \eqref{solutionI}. The existence of this lower end
threshold is clearly of importance in view of experimental realization but, as
being directly related to attenuation, it depends on the precise physical
context. Let us also remark that the output amplitude is sensitive to the input
amplitude values (in both regimes). In particular the curves (and dots) of
fig.\ref{schroedhyst} show that this dependence is roughly linear as the slopes
of the functions $B(b)$ in regimes I and III have approximately the same
slopes.

As a matter of fact we display on the fig.\ref{schroedevol} the evolution of
the amplitudes $|\psi_{0}(z)|=|\psi_{N}(z)|$  along the driving waveguides and
the amplitude in the middle of the array  $|\psi_{N/2}(z)|$ obtained from a
numerical simulation. The difference between the two output signals at $y=L/2$
in these two regimes differs substantially (approximately by two orders of
magnitude). 

The existence of bistable regimes that switch from a low to a high output is a
candidate for an ultra sensitive bistable light detector. Moreover such a
property holds also in the fully discrete case as displayed in
fig.\ref{fig:map} (though we do not have in that case an analytical
description). Particularly, if one injects in the lateral waveguides beams of
amplitude slightly below the supratransmission threshold $b_s$, a regime of
type III establishes (with small output amplitude). Then any perturbation of
the array by means of a weak input signal  destabilizes the system which, after
transient emission of gap solitons, reaches a type I  regime with large output
amplitude. This is precisely the process described in the intensity plot of
fig.\ref{fig:map} where the signal producing the switch carries 0.03\% of the
driving amplitude.

Note that in such a fully discrete case, the device works as a digital
detector, triggered by any signal received by the central waveguides. However
in the continuous case, the output signal amplitude can be (approximately)
linearly related to the input amplitude and, if the variable parameter is the
input amplitudes in the lateral waveguides, then the device can be operated as
an amplificator.

Acknowledgements: R. Kh. acknowledges financial support from NATO and thanks
International Centre for Theoretical Physics (Trieste, Italy) for the
hospitality.

\end{document}